\documentclass[aps,prl,twocolumn,psfig,showpacs]{revtex4}
\usepackage{amsfonts}
\usepackage{amsmath}
\usepackage{graphicx}
\usepackage{bm}
\usepackage{amssymb}
\usepackage{times}
\usepackage{dcolumn}
\usepackage{cases}
\usepackage{txfonts}
\usepackage{color}

\RequirePackage[hyperindex,colorlinks,bookmarksnumbered,plainpages=true]{hyperref}
\hypersetup{colorlinks,linkcolor=blue,urlcolor=blue,citecolor=blue}
\usepackage{hyperref}

\usepackage{ulem}%only for the command \sout = scrap
\renewcommand{\emph}[1]{\textit{#1}}%needed, if package  "ulem" is used

\definecolor{darkgreen}{rgb}{0,0.5,0}
\definecolor{darkblue}{rgb}{0,0,0.5}
\definecolor{darkred}{rgb}{.7,0,0}
\definecolor{purple}{rgb}{0.35,0,0.35}
\definecolor{orange}{rgb}{1,0.5,0}
\definecolor{grey}{rgb}{.6,.6,.6}

\newcommand{\Fig}[1]{Fig.~\ref{#1}}

\begin{document}

\title{Topology-driven phase transitions in the classical monomer-dimer-loop model}
\author{Sazi Li$^{1}$}
\author{Wei Li$^{2,1}$}
\email{w.li@physik.lmu.de}
\author{Ziyu Chen$^{1,3}$}
\email{chenzy@buaa.edu.cn}
\affiliation{$^{1}$ Department of Physics, Beihang University, Beijing 100191, China  \linebreak
$^{2}$ Physics Department, Arnold Sommerfeld Center for Theoretical Physics, and Center for NanoScience, Ludwig-Maximilians-Universit\"at, 80333 Munich, Germany
\linebreak
$^{3}$ Key Laboratory of Micro-nano Measurement-Manipulation and Physics (Ministry of Education), Beihang University, Beijing 100191, China}

\begin{abstract}
In this work, we investigate the classical loop models doped with monomers and dimers on a square lattice, whose partition function can be expressed as a tensor network (TN). In the thermodynamic limit, we use the boundary matrix product state technique to contract the partition function TN, and determine the thermodynamic properties with high accuracy. In this monomer-dimer-loop model, we find a second-order phase transition between a trivial monomer-condensation and a loop-condensation (LC) phases, which can not be distinguished by any local order parameter, while nevertheless the two phases have distinct topological properties. In the LC phase, we find two degenerate dominating eigenvalues in the transfer-matrix spectrum, as well as a non-vanishing (nonlocal) string order parameter, both of which identify the \textit{topological ergodicity breaking} in the LC phase and can serve as the order parameter for detecting the phase transitions. \end{abstract}

\date{\today}
\maketitle

\textit{Introduction.---}
Two dimensional (2D) monomer-dimer model has a quite venerable history in statistical mechanics \cite{Fowler-1937, Fisher-1961, Kasteleyn-1961, Fisher-1963}. The monomer-dimer model can be used to describe the absorption of molecules on the surface: the molecule can occupy two nearest neighboring sites and form a dimer, while the empty site is regarded as a monomer \cite{Fowler-1937}. The monomer-dimer model can also be related to other statistical models like Ising and height models \cite{Fisher-1961, Blote}, etc, thus it plays the role as a quite fundamental statistical model. On a square lattice, the fully packed dimer model is found to possess algebraic decaying dimer-dimer correlation, however, doping the system with monomers will drive the system out of the criticality and no phase transition occurs in a non-interacting monomer-dimer model at finite temperatures \cite{Alet-2005, Alet-2006, Li-2014}. On the other hand, if one introduces pairing interactions between the dimers, there exist phase transitions between the low-$T$ ordered phase and high-$T$ disordered one (Kosterlitz-Thouless type for fully packed case \cite{Alet-2005}, and second-order after monomer doping \cite{Alet-2006, Li-2014}).

Loop models are also widely studied in statistical mechanics, which is relevant for realistic physical systems and also constitutes a quite fundamental mathematical problem \cite{Fendley-2006, Nienhuis-1987}. The loop structure also plays an important role in certain quantum cases, like in the ground state of toric code \cite{Kitaev-2003}, the string-net model \cite{Levin-2004}, and the resonating Affleck-Kennedy-Lieb-Tasaki loop spin liquid states \cite{RAL}, etc. In Ref. \onlinecite{Castelnovo-2007}, Castelnovo and Chamon couple the toric code model to a thermal bath, and consider the thermal superposition of all possible loop coverings. They found that the concept of topological order also applies in this classical loop system, where the low-energy phase space decompose into several distinct topological sectors. The existence of distinct topological sectors breaks ergodicity. One needs to create/annihilate a loop with length propositional to system size, which has huge energy cost and rare probability to happen, in order to tunnel from one sector to another, it thus leads to the topological glass behavior \cite{Ng-2009}. The notion of topological entropy can be generalized to detect such nontrivial topological order in classical systems, by noticing that the topological constraint would also reduce the entropy in the classical case \cite{Castelnovo-2007}. Recently, Hermanns and Trebst have generalized this entropy characterization to general classical string-nets and verify that there are corresponding universal topological corrections in the Renyi entropy for a number of SU(N)$_k$ anyonic theories \cite{Hermanns-2014}.

In this work, we combine the two classical models and introduce a monomer-dimer-loop (MDL) model on a square lattice. The MDL model has a rather compact tensor network (TN) representation with a small bond dimension ($D=3$), and is thus amenable to high precision TN numerical simulations. TN-based numerical methods have been widely used to tackle statistical models and have been proved to be a very accurate and reliable tool \cite{TRG, SRG, Li-2010, Li-2014}. Through the TN numerical simulations, we show that the MDL model has a trivial disordered phase and a topologically ordered loop-condensation (LC) phase, with a second-order transition separating them. In addition, we characterize the LC phase with the vanishing gap of transfer-matrix spectrum and a nonlocal string order parameter (SOP), both of which can be used to pinpoint the phase transition.

\begin{figure}[htpb]
  \begin{center}
	\includegraphics[width=0.8\columnwidth]{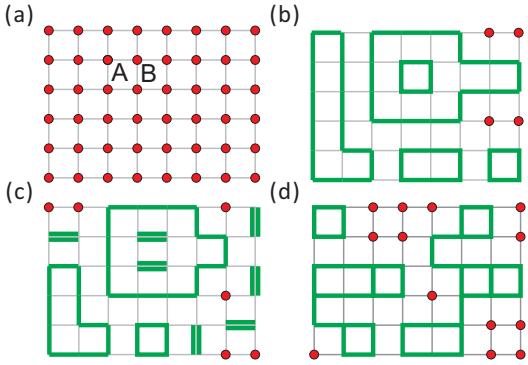}
  \end{center}
  \caption{(Color online) Snapshots of the classical phases of the MDL model on a square lattice: (a) The trivial monomer-condensation, (red) dots are the monomers, $A,B$ label the two sub-lattices; (b, c) Loop condensation in the monomer-loop and monomer-dimer-loop cases, respectively; (d) Loop condensation in the branching monomer-loop model (i.e., classical string-nets).}
  \label{illustra}
\end{figure}

\textit{Model and method.---}
Snapshots of several classical configurations in different phases of MDL model are shown in \Fig{illustra}. Summing over all possible classical configurations, we have the partition function
\begin{equation}
\Xi = \sum_{\{c\}} \exp{[-\beta ( \mu N_m + \nu N_b + u N_d)]},
\label{eq-GPF}
\end{equation}
where $\{c\}$ means the set of all classical monomer-dimer-loop configurations, $N_{m}$ is the total number of vertices occupied by a monomer, which has an energy of $\mu$; $N_b$ counts the number of edges occupied by a loop, and $\nu$ is the energy per bond of a loop; $N_{d}$ is the total number of vertices linked by a dimer (with energy per dimer as $2u$). In the following, $\nu=1$ is set as the energy scale if not otherwise specified.

\begin{figure}
\centering
\includegraphics[width=0.47\textwidth]{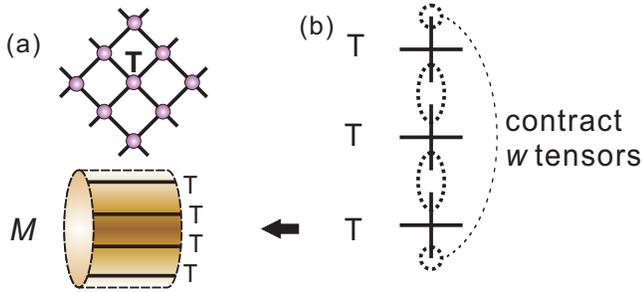}
\caption{(Color online) (a) TN representation of the partition function, on a $\pi/4$ tilted square lattice. (b) The construction of the transfer matrix $M$ on a cylindrical geometry by contracting a column of ($w$) rank-4 tensors. The dashed lines denote the contractions between two tensors.}
\label{TN+M}
\end{figure}

The partition function of the monomer-loop model has a simple TN representation which form a $\pi/4$ tilted square lattice, which represents the partition function $Z$, as shown in \Fig{TN+M}(a). The partition function TN consists of tensors $T_{s_1,s_2,s_3,s_4}$ located at each vertex, which has four indices $s_i$ ($i \in \{1,2,3,4\}$) corresponding to the four geometric bonds. Each index has a finite bond dimension ($D=3$), i.e., $s_i \in \{0,1,2\}$: $s_i=0(1)$ means the absence (presence) of a loop bond, $s_i=2$ represents the presence of a dimer, on the specific edge $s_i$. We properly initiate the tensor $T$, to make sure that a specific lattice site is either occupied by a loop [$T_{1,1,0,0}=T_{1,0,1,0}=T_{1,0,0,1}=T_{0,1,1,0}=T_{0,1,0,1}=T_{0,0,1,1}=\exp{(-\nu/T)}$], a monomer [$T_{0,0,0,0}=\exp{(-\mu/T)}$], or by a dimer [$T_{2,0,0,0}=T_{0,2,0,0}=T_{0,0,2,0}=T_{0,0,0,2}=\exp{(-u/T)}$], and the rest elements are zero (forbidden).

To calculate the thermodynamics of the MDL model, one has to efficiently (and accurately) contract the partition function TN for calculating free energy per site $f$, energy per site $e$ and other thermodynamic quantities. In this work, we define the system on two kinds of geometries: for the infinitely large 2D lattice system, we adopt the iTEBD method \cite{Vidal2007, Orus} for accurate contractions; for the cylindrical geometry with finite (small) circumferences and infinite length, we diagonalize the spectrum of the transfer matrix and evaluate properties with its dominating eigensystems. iTEBD was initially proposed for efficient simulations of the time evolution and the ground state property (through imaginary-time evolution) of 1D quantum systems, and then generalized to calculate the thermodynamics of 2D classical statistical models \cite{Orus} and also 1D quantum lattice models \cite{LTRG}. In our practical simulations, we perform the contraction of MPS with transfer MPO until the prescribed convergence criterion is reached, say, free energy per site converges to $10^{-14}$ (almost machine precision). The total number of iterations is around $10^{3\sim4}$, depending on the temperatures and the physical parameters of the model. The retained bond dimension of the boundary MPS $D_c \approx 100$, the convergence with $D_c$ is always checked,  the truncation error is less than $10^{-6}$ at the critical point, and reaches the machine precision away from the critical points.

\textit{Monomer-Loop model.---} In the partition function Eq. (\ref{eq-GPF}), if we forbid the dimer occupation (i.e., $u \to \infty$), the model is reduced to a monomer-loop model, which can be related to the well-known Ising model. For instance, the triangular lattice Ising model can be mapped to a monomer-loop model on its dual honeycomb lattice, where the loops are the magnetic domain walls separating spins which have opposite orientations, and the monomers are the topological excitations on top of that \cite{Ng-2009}. In our present model, we treat directly the monomer-loop picture, and thus can tune monomer energy $\mu$ continuously (with fixed $\nu=1$), making the simulations beyond the exact Ising model mapping (also notice that we are simulating the loop models on a square lattice, and do not allow loop crossovers, which thus in the beginning lacks an exact Ising mapping).

\begin{figure}
\centering
\includegraphics[width=0.48\textwidth]{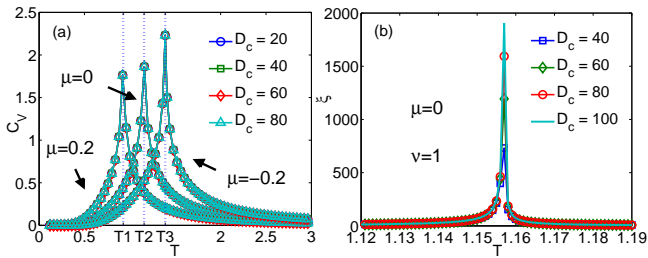}
\caption{(Color online) (a) The specific heat $C_V$ of  the monomer-loop model with $\mu=0, \pm0.2$ and $\nu=1$, where $T_c$'s correspond to $0.927, 1.157, 1.39$ respectively. (b) The correlation length $\xi$ versus temperature $T$ of the classical loop model. The heights of the peaks at $T_c$ grow by increasing $D_c$.}
\label{02-Cv+xi}
\end{figure}

Firstly, we investigate the monomer-loop model with $\mu=0, \pm0.2$. The specific heat $C_V$ curve is shown in \Fig{02-Cv+xi}(a), which is computed by taking first-order derivative (versus $T$) of the energy per site. The latter is calculated via $e$ = $Z^*/Z$, where $Z^*$ is obtained by contracting the TN with one $T$ tensor replaced with an impurity tensor $T^I = \mu T_{0,0,0,0} + \nu (T_{1,1,0,0} + T_{1,0,1,0} + T_{1,0,0,1} + T_{0,1,1,0} + T_{0,1,0,1} + T_{0,0,1,1} )$. In \Fig{02-Cv+xi}(a), divergent peaks appear at $T_c \approx 0.927, 1.157, 1.39$, for $\mu=0.2, 0, -0.2$, respectively, suggesting the presence of second-order phase transitions.

In addition, to confirm the existence of the phase transition, we calculate the correlation length $\xi$ of the monomer-loop model via the following formula
\begin{equation}
\xi = 1/\ln (\frac{\lambda_{1}}{\lambda_{2}}),
\label{eq-xi}
\end{equation}
where $\lambda_{1}$ ($\lambda_{2}$) is the largest (second-largest) eigenvalue of the transfer matrix $M$ (in case the largest eigenvalues are $n$-fold degenerate, $\lambda_2$ is the $n+1$ largest one). In \Fig{02-Cv+xi}(b), we observe that the correlation length $\xi$ also shows a divergent peak at $T_{c}$, confirming the criticality at the transition point.

\begin{figure}
\centering
\includegraphics[width=0.48\textwidth]{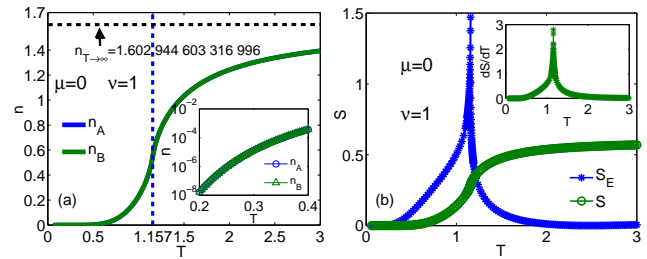}
\caption{(Color online) (a) The mean bond occupation number $n_A$ and $n_B$ on the vertices A and B in the monomer-loop model ($\mu=0, \nu=1$), the horizontal dashed line shows the bond density in the $T=\infty$ limit. The inset amplifies and shows the low temperature behaviors. (b) The entropy $S$ and the entanglement entropy $S_E$ of the classical loop model with $\mu=0, \nu=1$. Inset shows the first-order derivative of entropy $dS/dT$.}
\label{02-n+S}
\end{figure}

However, interestingly, we find no local order parameters for detecting this phase transition, since both the high- and low-$T$ phases are disordered and have no symmetry breaking. We show the numerical results of bond density ($n_A$ and $n_B$)  in Fig. \ref{02-n+S} (a), which counts the average bonds (of the loops) per site. The results (with $\mu=0$) are shown in Fig. \ref{02-n+S}(a), from which we can see that the low-$T$ phase has relatively low bond density and thus can be regarded as monomer-condensation (MC), while the high-$T$ region is a loop-condensation (LC) phase. Although $n_A$ and $n_B$ change from zero to nonzero values when $T$ increases, they change smoothly through the transition point. In addition, the same bond densities $n_A = n_B$ are observed for all temperatures $T$, which suggests that the symmetry between two sub-lattices is also intact. Therefore, we conclude that the bond density $n$ can not serve as a local order parameter for distinguishing two phases. Besides, in \Fig{02-n+S}(a), we also show the bond density $n=1.602\, 944\, 603\, 316\, 996$ (with 16 converged significance digits) in the $T=\infty$ limit, where the state is an equal-weight (classical) superposition of all possible monomer-loop configurations. Compared to the dimer density $n_d=0.638\, 123\, 109\, 228\, 547$ in monomer-dimer model \cite{Li-2014}, we find that $n>2n_d$ here.

We also investigate the entropy of the system, including two kinds of entropies, i.e., the conventional thermodynamic entropy $S = (U-F)/T$ and the formal ``entanglement entropy" $S_{\rm{E}}$ evaluated from the boundary MPS.  The latter can be obtained by $S_{\rm{E}} = -\sum_i \Lambda_i^2 \ln(\Lambda_i^2)$, where $\Lambda_i$'s are the Schimidt spectrum of the decomposition on a bond $i$. As shown in \Fig{02-n+S}(b), the bipartite entanglement entropy $S_{\rm{E}}$ of the classical loop model shows a clearly divergent peak at $T_c$, indicating the occurrence of a phase transition. On the contrary, the conventional thermodynamic entropy $S$ is smooth around $T_c$. However, its first-order derivative has a divergent peak, as shown in the inset of \Fig{02-n+S}(b), which is not surprising since $\frac{\partial S}{\partial T} = \frac{C_{V}}{T}$. Therefore, we see that this ``entanglement entropy" $S_{\rm{E}}$ is more sensitive to the phase transition (than the thermodynamic entropy $S$), and serves as a very useful numerical tool detecting phase transitions. Similar behaviors have already been seen in our previous tensor-network study of the monomer-dimer model \cite{Li-2014}.

In order to further investigate the phase transitions, we also define the MDL model on cylindrical geometries. On the cylinders with finite circumferences (and infinite length), we can contract the TN exactly and thus obtain the thermodynamic quantities. Specifically, we start from both ends, and contract the boundary vectors with the transfer matrix $M$ consisting of a column of rank-4 tensors [see \Fig{TN+M}(b)]. Repeating the contraction process until both boundary vectors converge, with which we can evaluate the observables like the energy expectation values. As the cylinder widths increase, the observables should eventually converge to the thermodynamic limit results obtained with iTEBD contractions above.

\begin{figure}
\centering
\includegraphics[width=0.5\textwidth]{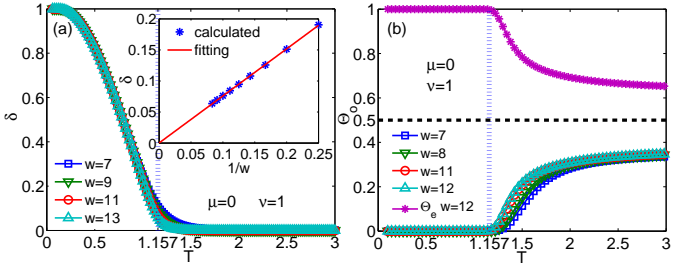}
\caption{(Color online)  Topological characterization of the monomer-loop model for various cylinder widths $w$. (a) The gap of the transfer matrix $\delta$ vanishes when $T>T_c$, the inset shows that the extrapolated $\delta\sim0$ in the $w=\infty$ limit. (b) The odd string order parameter $\Theta_{o}$ is zero when $T\leq T_c$ and nonzero $T>T_c$; $\Theta_{e}$ also changes its behavior at $T_c$.}
\label{02-topo}
\end{figure}

We calculate the (normalized) gap of the transfer matrix $\delta=|\frac{\lambda_{1}-\lambda_{2}}{\lambda_{1}}|$ for various cylinder widths $w$, which is shown in \Fig{02-topo}(a). In the $T>T_c$ region, we observe two degenerate dominating eigenstates in the spectrum of transfer-matrix $M$. In particular, as shown in the inset of \Fig{02-topo}(a), $\delta$ extrapolates to zero at critical point $T=T_c$, in the thermodynamic limit. Therefore, $\delta$ can be taken as an order parameter detecting the phase transition between the low- and high-$T$ phases.

We are also interested in the parity of the dominating eigenvector $\chi$ of transfer matrix $M$. Since the loops are closed in the MDL model, $M$ conserves the parity symmetry. When the cylinder is cut into two halves vertically, the number of intersected bonds by the cut is either even or odd, which defines the parity of eigenvector $\chi$. For cylinders with open ends (i.e., no dangling bonds on the edges), all the allowed configurations constitute the even sector, and the dominating eigenvector $\chi_e$ in this sector is with even parity. On the other hand, if we introduce odd number of open strings on the cylinder, stretching from the very left boundary to the rightmost side, then all the allowed configurations constitute an odd sector, with dominating eigenvector $\chi_o$ of odd parity. \Fig{02-topo}(a) shows that the dominating even and odd eigenvectors ($\chi_e$ and $\chi_o$) become degenerate when $T>T_c$.

Furthermore, we introduce a string operator $\Theta$ winding around the cylinder to measure the parity of eigenvectors $\chi$. The string operator $\Theta = \prod_{i=1}^w P_i$ is a product of the operator
\begin{equation}
P=
\left(
  \begin{array}{cc}
     1 & 0\\
    0 & -1\\
  \end{array}
\right)
\end{equation}
which lives on the horizontal edge \cite{SM}. $P=1 (-1)$ if the edge is not occupied (occupied by a bond). Therefore, the expectation value of the product of $P$ tells whether the system is in the even or odd sector. In \Fig{02-topo}(b), we thread an open string in the cylinder, and show the numerical results of $\Theta_{o} = |\chi_{o} \Theta M \chi^{*}_{o} /( \chi_{o} M \chi^{*}_{o} + \chi_{e} M \chi^{*}_{e} )|$ for various cylinder widths, where the partition function $\chi_{e}$($\chi_{o}$) is the even(odd) dominating eigenvectors. We observe that $\Theta_o$ is a constant zero for $T<T_c$, and becomes nonzero when $T>T_c$. In the meantime, we also show the even $\Theta_{e} = |\chi_{e} \Theta M \chi^{*}_{e} /( \chi_{o} M \chi^{*}_{o} + \chi_{e} M \chi^{*}_{e} )|$, which is a constant one in the trivial MC phase, while also changes its behavior at $T_c$. Thus, $\Theta_{o}$ can also be taken as an order parameter for detecting the phase transition, called nonlocal string order parameter.

\begin{figure}
\centering
\includegraphics[width=0.48\textwidth]{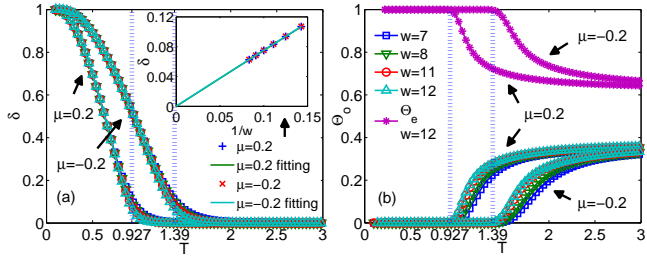}
\caption{(Color online) The topological properties of the monomer-loop model with $\mu=\pm0.2$ and $\nu=1$ for various cylinder widths $w$. (a) The gap of transfer matrix $\delta$ vanishes when $T>T_c$, the inset shows $\delta$ extrapolates to zero at $T_c$. (b) The string order parameters $\Theta_{o}$ and $\Theta_{e}$.}
\label{02-mu-topo}
\end{figure}

\begin{figure}
\centering
\includegraphics[width=0.45\textwidth]{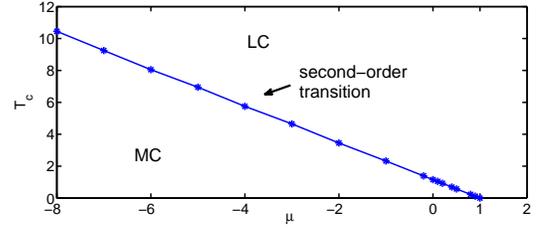}
\caption{(Color online) The phase diagram $\mu-T$ of the monomer-loop model. The disordered monomer-condensation (MC) and topologically ordered loop-condensation (LC) phases are separated by a second-order transition line when $\mu \in (-\infty, 1]$.}
\label{phase}
\end{figure}

In addition to $\mu=0$ case, \Fig{02-mu-topo} shows the corresponding results of the monomer-loop model with $\mu=\pm0.2$, including the (normalized) gap $\delta$ in \Fig{02-mu-topo}(a) and SOP in \Fig{02-mu-topo}(b). Similar behaviors can be seen as in the $\mu=0$ case. In \Fig{phase}, we tune the monomer doping parameters $\mu$, collect the phase transition points of the monomer-loop model with various parameters $\mu$, and obtain the $\mu-T$ phase diagram. When $\mu<1$, there exist second-order phase transitions separating the low- and high-$T$ phases. $T_c$ decreases with increasing monomer energy until $\mu =1$, where $T_c=0$, i.e., no phase transitions.

Here we would like to address some remarks on the classical topological order in the monomer-loop model. As a consequence of the loop condensation [\Fig{illustra}(b)], there exist two degenerate eigenvectors $\chi$ in the high-$T$ phase, meaning that the phase space is decomposed into two distinct topological sectors, which leads to a topological ergodicity breaking. One gets exactly the same results by evaluating the thermodynamic quantities in either sector, while it is not possible to shift from one sector to the other by changing the loop configurations only locally. This glassy behavior of the LC phase is due to topological reasons, therefore the LC phase can also be called a topological glass \cite{Ng-2009}, and the phase transition between MC and LC phases is thus a topology-driven transition.

%In particular, the odd sector of LC phase, which can be regarded to possess odd number of open strings of length scale as large as the system size, has a non-negligible weight in the overall partition function; however, this is not the case in the MC phase, since there are few loops ($n\approx0$) and the existence of an open string of macroscopic length is highly unlikely [\Fig{illustra}(a)]. As shown in \Fig{02-topo}(b), by threading an open string on the cylinder, we can see clearly different behaviors between the LC and the trivial MC phases, the latter has no ``coupling" to this ``flux" ($\Theta_o$=0 for $T<T_c$), while the former shows nonzero $\Theta_o$ response.

\textit{Monomer-Dimer-Loop model.---}
In the monomer-loop model, the dimer occupation was not allowed (i.e., effectively $u=\infty$ in Eq. \ref{eq-GPF}). The dimer can be regarded as the ``minimal" loop of length two [shown in \Fig{illustra} (c)]. In the following, we switch on dimer coverings, and study the full MDL model with $u=3$ (and $\mu=0, \nu=1$).

As in the monomer-loop model case, we also calculate the specific heat $C_V$, the correlation length $\xi$, bond density $n$ and the entropy (the thermodynamic entropy $S$ and the entanglement entropy $S_{\rm{E}}$) of the system \cite{SM}. $C_V$, $\xi$, $S_{\rm{E}}$ and the first-order derivative of the thermodynamic entropy all show a divergent peak at $T_c \approx 1.179$, indicating the occurrence of a second-order phase transition. On the other hand, the bond density (the presence of a dimer is considered to be equivalent to two overlapping bonds) is smooth around $T_c$ and $n_A=n_B$. Thus the bond density again can not serve as a local order parameter.

\begin{figure}
\centering
\includegraphics[width=0.5\textwidth]{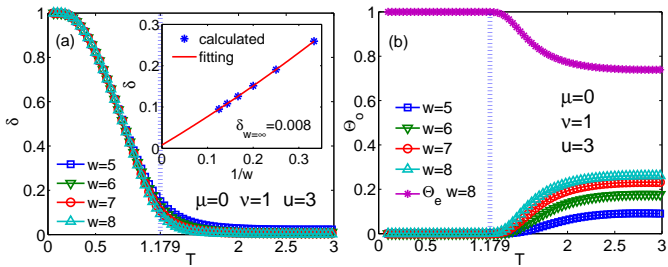}
\caption{(Color online) The topological properties of the MDL model with $\mu=0,\nu=1,u=3$ for various cylinder widths $w$. (a) The gap of transfer matrix $\delta$ serves as an order parameter; Inset: $\delta$ at $T_c$ for various cylinder widths and their extrapolation. (b) The string order parameter $\Theta_{o}$ and $\Theta_{e}$.}
\label{022-topo}
\end{figure}

The normalized gap $\delta$ and SOP $\Theta_o$ in the MDL model are shown in \Fig{022-topo}. In \Fig{022-topo}(a), we again see a twofold degeneracy in the transfer matrix spectrum in the LC phase, indicating that the LC phase is also topologically ordered in the MDL model. The inset of \Fig{022-topo}(a) shows $\delta$ extrapolates to zero (in the thermodynamic limit) at $T_c$. In \Fig{022-topo}(b), $\Theta_{o,e}$ are shown, in which $\Theta_o$ is nonvanishing in the LC phase, suggesting that it can also identify the classical topological order in the MDL model. In summary, similar to the monomer-loop case, the classical topological order exists in the LC phase of the general MDL model and the second-order phase transition separates the trivial disordered MC and topologically ordered LC phases. $\delta$ and $\Theta_o$ can be used as order parameters to characterize the topology-driven classical phase transition well.

\textit{Conclusion and outlook.---}
In this work, we have systematically studied the classical loop model with monomer and dimer doping. Using the boundary MPS contraction method, we evaluate the partition function TN and obtain the thermodynamic properties including the specific heat $C_{V}$, correlation length $\xi$ and entropies. There exist second-order phase transitions separating the trivial monomer-condensation and the loop-condensation phases, which can not be described by the local order parameters like bond density $n$. However, in the LC phase, we find twofold degenerate dominating eigenvalues in the transfer matrix spectrum, one in even and the other in odd topological sectors. The existence of two topological sectors actually breaks the ergodicity. The non-vanishing nonlocal order parameter SOP $\Theta_o$ can also be used to distinguish two sectors and thus detect the phase transition. Therefore, these two phases can be identified by their distinct topological properties, and the phase transition between them belongs to a topology-driven type.

Besides the closed loop cases studied in the MDL model above, it is also interesting to consider the model with the branching loops [see \Fig{illustra}(d)], i.e., the classical string-net model. It is quite straightforward to generalize the tensor-network representation here to the classical string-nets, and our preliminary calculations show that there also exists a second-order phase transition between LC and MC phases. However, owing to the existence of branching loops, the transfer-matrix breaks the parity symmetry and no longer has the well-defined even and odd topological sectors as the MDL model has here. Our study of the classical string-nets will be published elsewhere.

\section{Acknowledgement}
W.L. acknowledges Hong-Hao Tu for helpful discussions. This work was supported in part by the National Natural Sciences Foundation of China (Grants No. 11274033, and No. 11474015), Major Program of Instrument of the National Natural Sciences Foundation of China (Grant No. 61227902), Sub Project No. XX973 (XX5XX), and the Research Fund for the Doctoral Program of Higher Education of China (Grant No. 20131102130005 ). W.L. further acknowledges support by the DFG through Grant No. SFB-TR12 and Cluster of Excellence NIM.

\onecolumngrid
\vspace{2cm}
\begin{center}
{\bf\large Supplementary materials for ``Topology-driven phase transitions in the classical monomer-dimer-loop model"}
\end{center}
\vspace{0.1cm}
\setcounter{equation}{0}
\renewcommand{\theequation}{S\arabic{equation}}
\setcounter{figure}{0}
\renewcommand{\thefigure}{S\arabic{figure}}
\section{Tensor network representation}
In this part, we introduce the partition function TN representation of these models and the contraction method of the nonlocal string order parameter $\Theta_{o,e}$.

\begin{figure}[htpb]
  \begin{center}
	\includegraphics[width=0.8\columnwidth]{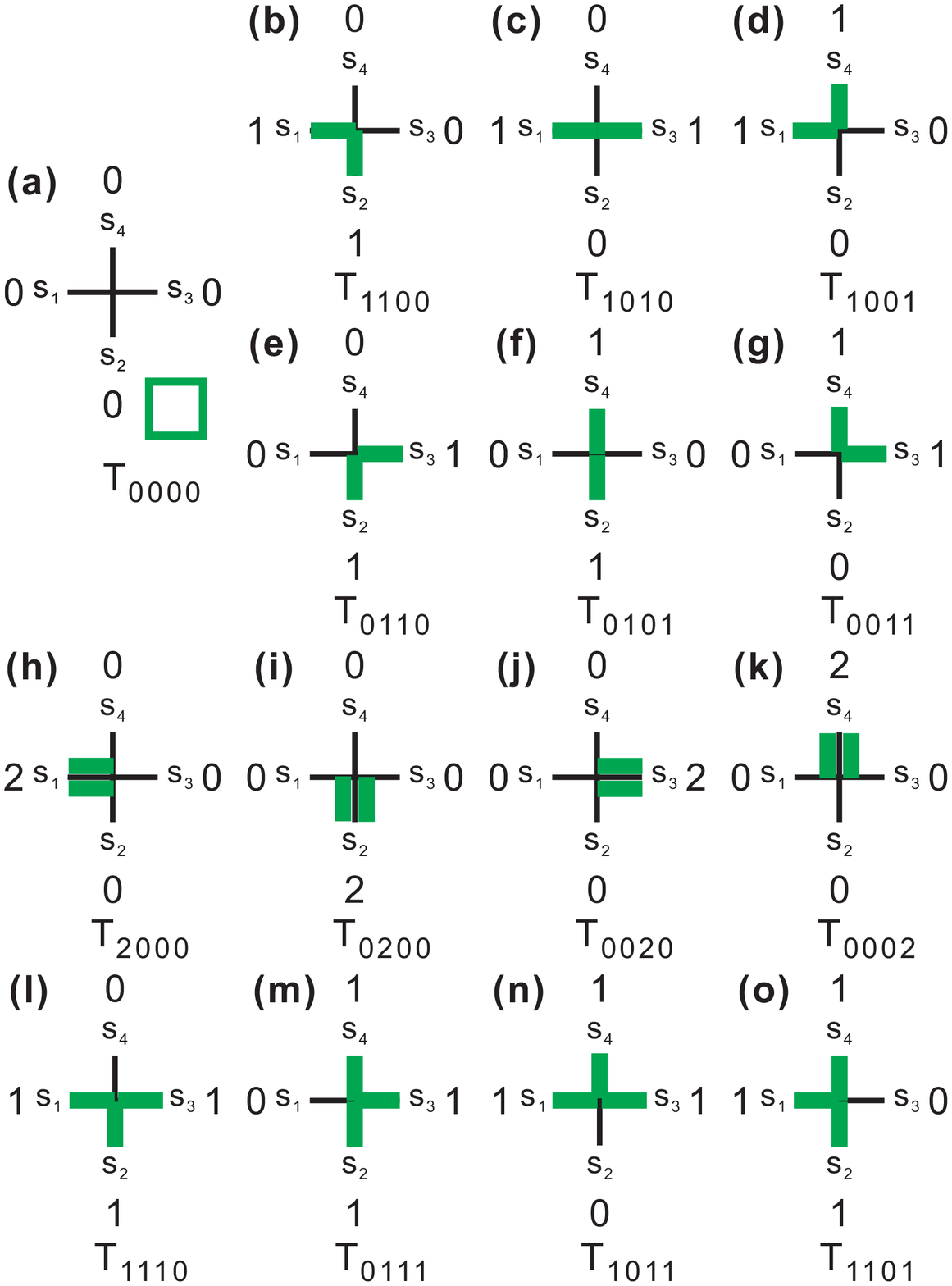}
  \end{center}
  \caption{(Color online) The eleven (allowed) nonzero elements of the vertex tensor $T_{s_1,s_2,s_3,s_4}$ in the partition function TN representation and their corresponding loop configurations. In the TN, a lattice site occupied by: (a) a monomer; (b-g) a loop; (h-k) a ``thick" dimer; (l-o) a branching loop.}
  \label{tensor}
\end{figure}

Since each vertex is allowed to be covered by at most one loop or a ``thick" dimer in the MDL model, there are eleven nonzero elements in the vertex tensor $T$. The allowed nonzero tensor elements $T_{s_1,s_2,s_3,s_4}$ and their corresponding classical loop configurations are schematically shown in \Fig{tensor} (a-k). $T_{0,0,0,0} = \exp(-\beta \mu)$ corresponds to the absence of any loop, i.e., a monomer [\Fig{tensor} (a)], $T_{1,1,0,0} = T_{1,0,1,0} = T_{1,0,0,1} = T_{0,1,1,0} = T_{0,1,0,1} = T_{0,0,1,1} = \exp(-\beta \nu) $ represent the one-loop configurations [\Fig{tensor} (b-g)], $T_{2,1,1,1} = T_{1,2,1,1} = T_{1,1,2,1} = T_{1,1,1,2} = \exp(-\beta u)$ describe the vertex with a dimer [\Fig{tensor} (h-k)], and $T_{1,1,1,0} = T_{0,1,1,1} = T_{1,0,1,1} = T_{1,1,0,1} = \exp(-\beta \nu)$ introduce the branching loop configurations [\Fig{tensor} (l-o)].

\begin{figure}[H]
\centering
	\includegraphics[width=0.8\columnwidth]{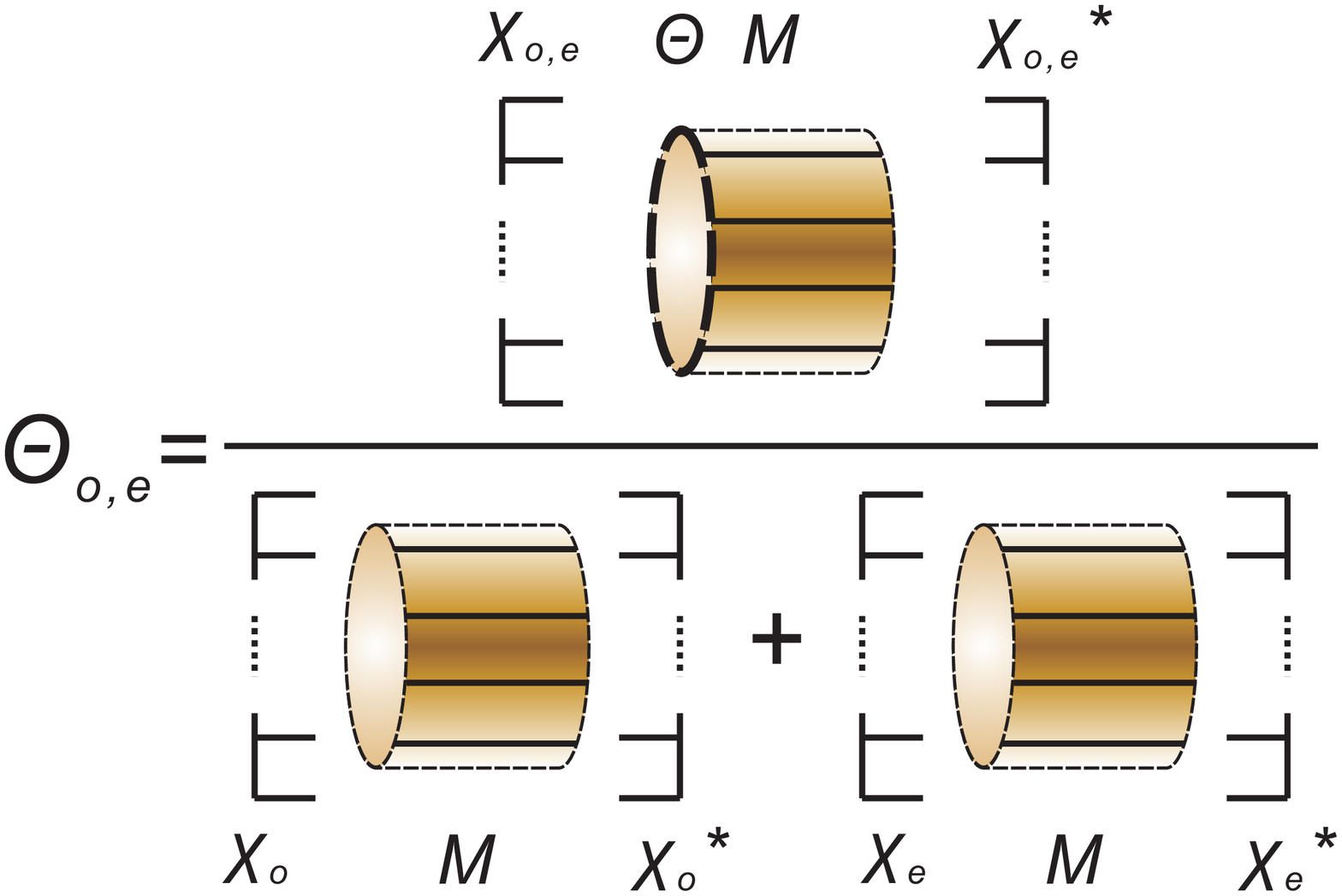}
  \caption{(Color online) The contraction method of the nonlocal string order parameter $\Theta_{o,e}$ on a cylindrical geometry, corresponding to the equations for $\Theta_{o,e}$ in the main text. The definition of $\chi_{o,e}, M, \Theta$ is also explained in the main text.}
  \label{theta}
\end{figure}

In \Fig{theta}, we show the way of evaluating $\Theta_{o,e}$ on a cylindrical geometry. Denominator is the partition function, and the numerator measures the expectation value of $\Theta$ in the odd(even) sector.

\begin{figure}[H]
\centering
\includegraphics[width=0.5\textwidth]{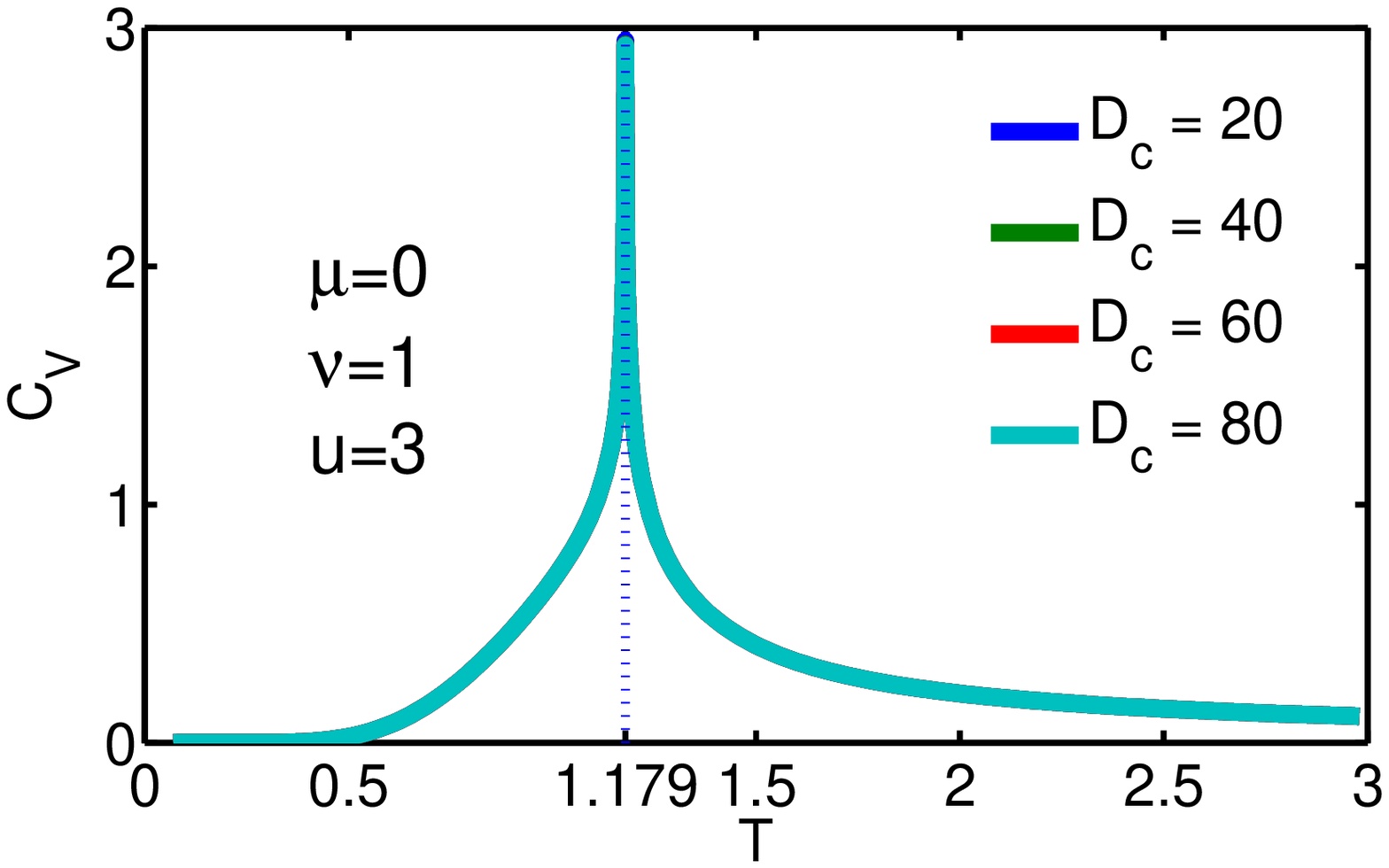}
\caption{(Color online) The specific heat $C_V$ of the MDL model with the energy $u=3$ of a dimer and fixed $\mu=0$, $\nu=1$.
\label{022-Cv}}
\end{figure}

\begin{figure}[H]
\centering
\includegraphics[width=0.5\textwidth]{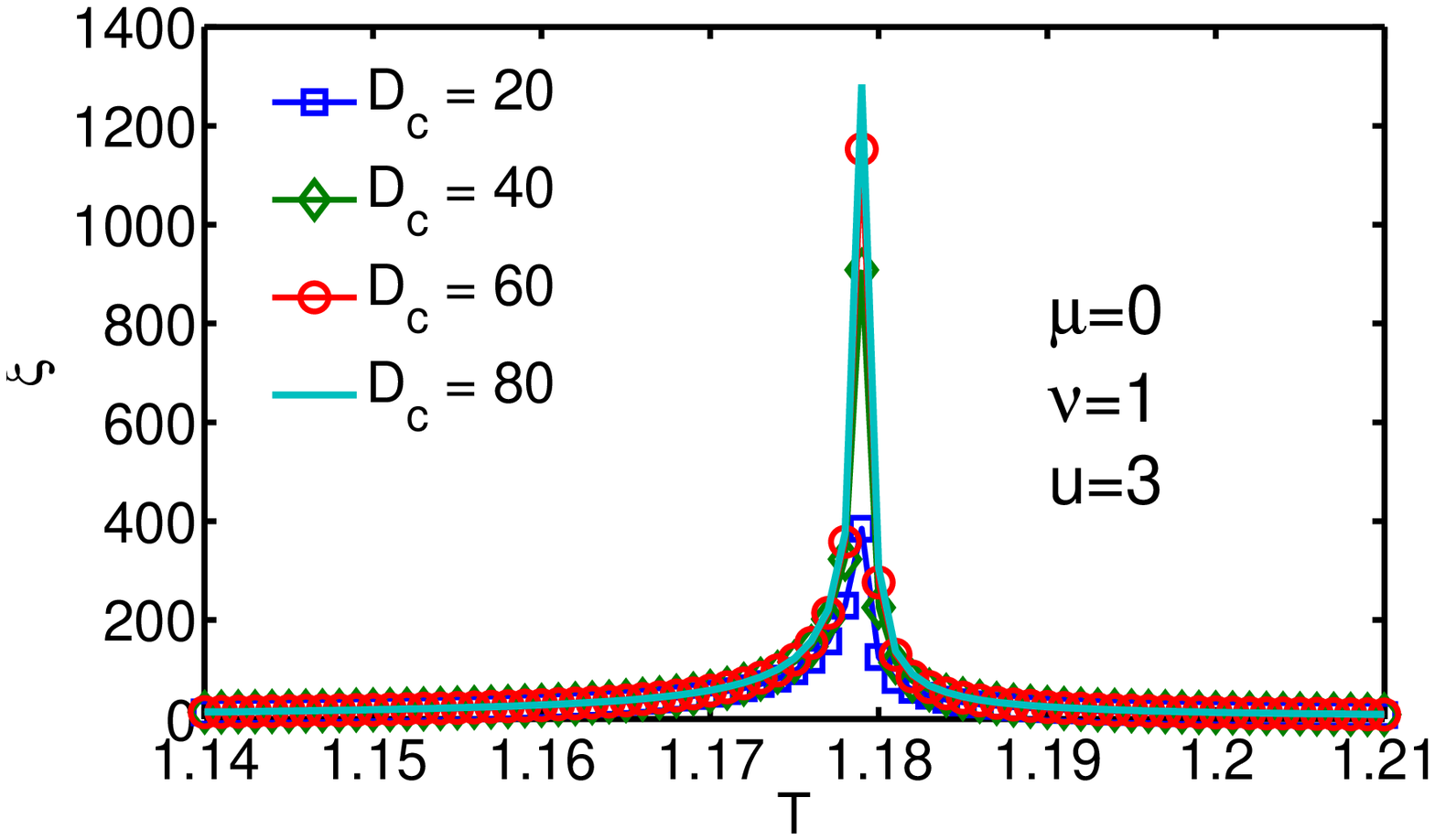}
\caption{(Color online) The correlation length of the MDL model with $\mu=0, \nu=1, u=3$. The heights of the peaks at $T_c$ grow with the increase of $D_c$.
\label{022-xi}}
\end{figure}

\begin{figure}[H]
\centering
\includegraphics[width=0.5\textwidth]{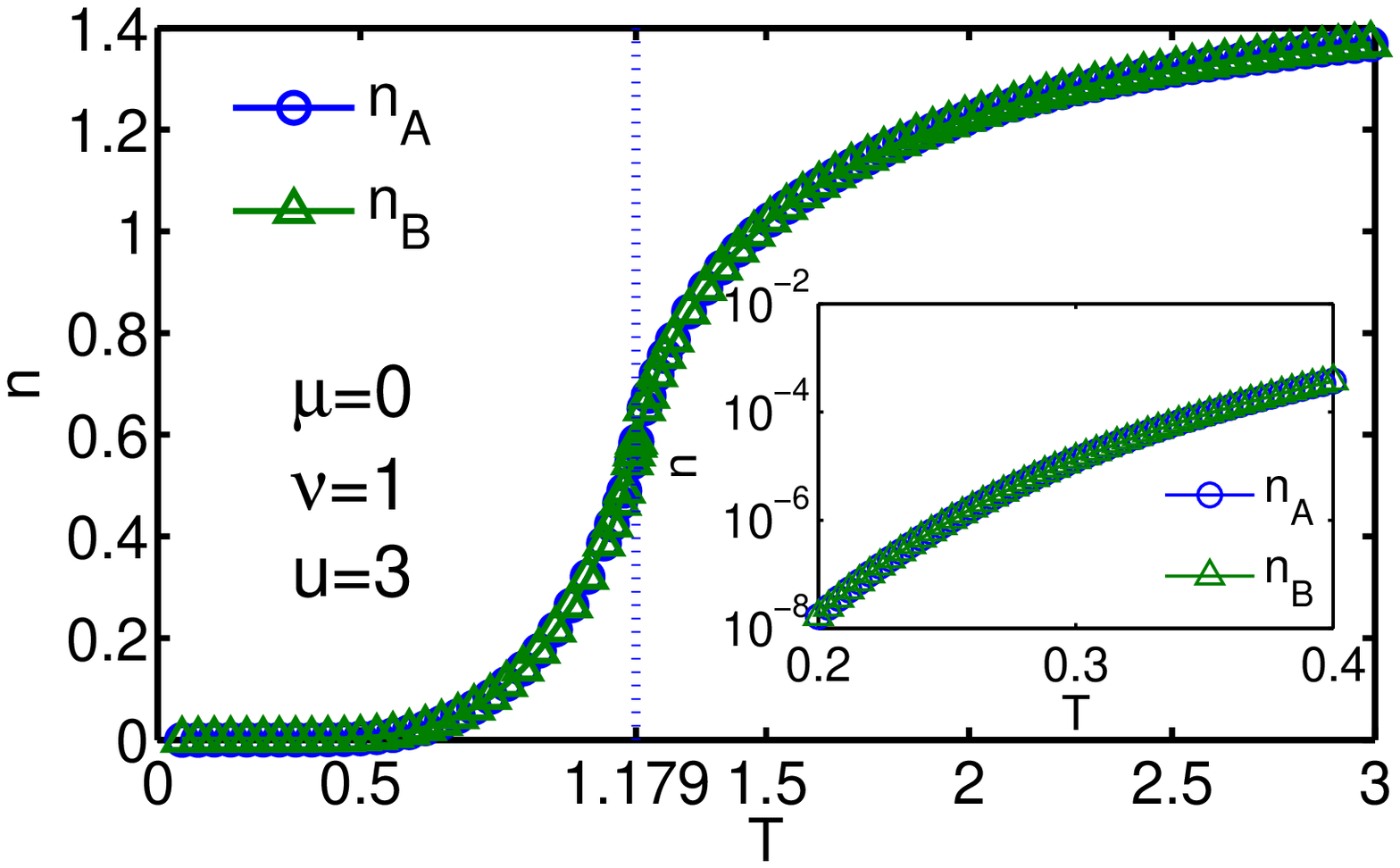}
\caption{(Color online) The bond density per site $n_A$ and $n_B$ of the MDL model with $\mu=0, \nu=1, u=3$. Inset: The amplification of $n_A$ and $n_B$ at low temperature region.
\label{022-n}}
\end{figure}

{\section{Specific heat,  correlation length, and entropy results of a monomer-dimer-loop model}}

For the MDL model with $\mu=0, \nu=1, u=3$, we also calculate its thermodynamic properties: the specific heat $C_V$, the correlation length $\xi$, the bond density $n$ and the entropy that includes the traditional thermodynamic entropy $S$ and the entanglement entropy $S_E$. In this part, we show the results as follows.

\Fig{022-Cv} shows our computed specific heat $C_V$ of the MDL model with the energy $u=3$ of a dimer and fixed $\mu=0, \nu=1$, where a divergent peak of $C_V$ occurs at $T_c\approx1.179$, uncovering the existence of the second-order phase transition.

In \Fig{022-xi}, we show the correlation length $\xi$ of the MDL model. From \Fig{022-xi}, we can also observe a divergent peak at $T_{c}$, the second-order phase transition point.

The bond densities per site $n_A$ and $n_B$ on vertexes A and B in the MDL model with $\mu=0, \nu=1, u=3$ are shown in \Fig{022-n}. Both $n_A$ and $n_B$ are changing smoothly around $T_c$ and $n_A$ is equal to $n_B$ for all temperatures, meaning that the symmetry between two sub-lattices in the system is not broken.

\begin{figure}[H]
\centering
\includegraphics[width=0.5\textwidth]{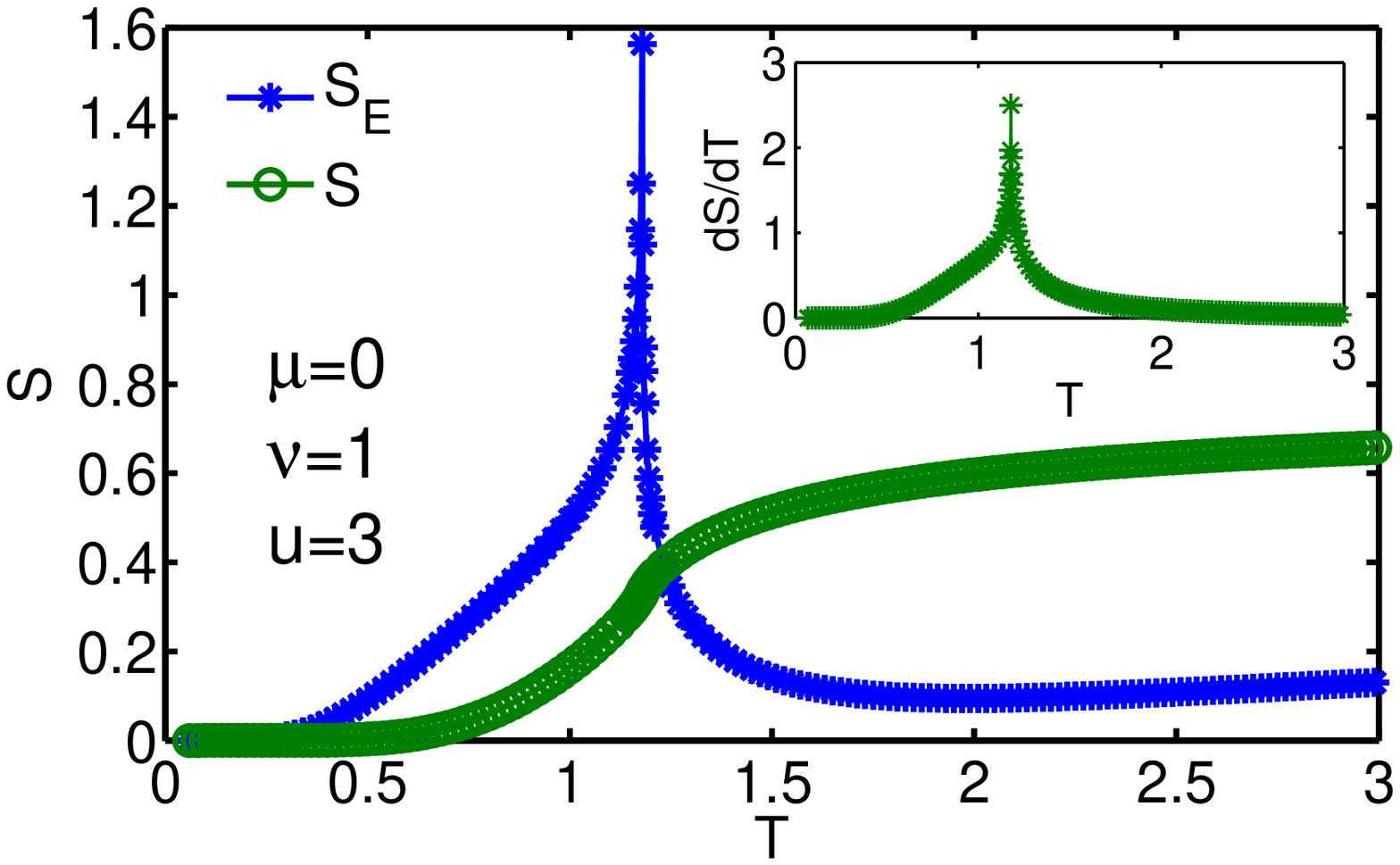}
\caption{(Color online) The thermodynamic entropy $S$ and the entanglement entropy $S_E$ of the MDL model with $\mu=0, \nu=1, u=3$. The inset shows the first-order derivative of entropy $S$ for temperature $dS/dT$.
\label{022-S}}
\end{figure}

The thermodynamic entropy $S$ and the entanglement entropy $S_E$ of the MDL model with $\mu=0, \nu=1, u=3$ are shown in \Fig{022-S}. $S_E$ shows a divergent peak at $T_{c}$, while $S$ is smooth around $T_c$, and its singularity can only be seen after taking a first-order derivative over $T$, which is shown in the inset of \Fig{022-S}.

\end{document}